\documentclass[amsmath,amssymb,aps,physrev,twocolumn,superscriptaddress]{revtex4-2}
\usepackage{graphicx}% Include figure files
\usepackage{dcolumn}% Align table columns on decimal point
\usepackage{bm}% bold math
\usepackage{graphicx}
\usepackage{xcolor}

\begin{document}

% Use the \preprint command to place your local institutional report
% number in the upper righthand corner of the title page in preprint mode.
% Multiple \preprint commands are allowed.
% Use the 'preprintnumbers' class option to override journal defaults
% to display numbers if necessary
%\preprint{}

%Title of paper
\title{Coexistence of superconductivity and topological band in a van der Waals $\bm{\mathrm{Sn}_{1-x}\mathrm{In}_{x}\mathrm{Bi_{2}Te_{4}}}$ crystal}

% repeat the \author .. \affiliation  etc. as needed
% \email, \thanks, \homepage, \altaffiliation all apply to the current
% author. Explanatory text should go in the []'s, actual e-mail
% address or url should go in the {}'s for \email and \homepage.
% Please use the appropriate macro foreach each type of information

% \affiliation command applies to all authors since the last
% \affiliation command. The \affiliation command should follow the
% other information
% \affiliation can be followed by \email, \homepage, \thanks as well.
%\author{Hoyeon Jeon}
%\email[]{Your e-mail address}
%\homepage[]{Your web page}
%\thanks{}
%\altaffiliation{}

\author{Hoyeon Jeon}
\author{Saban Hus}
\author{Jewook Park}
\affiliation{Center for Nanophase Materials Sciences, Oak Ridge National Laboratory, Oak Ridge, Tennessee 37831, USA}
\author{Qiangsheng Lu}
\author{Seoung-Hun Kang}
\author{Mina Yoon}
\author{Robert G. Moore}
\author{Jiaqiang Yan}
\author{Michael A. McGuire}
\affiliation{Materials Science and Technology Division, Oak Ridge National Laboratory, Oak Ridge, Tennessee 37831, USA}
\author{An-Ping Li*}
\affiliation{Center for Nanophase Materials Sciences, Oak Ridge National Laboratory, Oak Ridge, Tennessee 37831, USA}
%Collaboration name if desired (requires use of superscriptaddress
%option in \documentclass). \noaffiliation is required (may also be
%used with the \author command).
%\collaboration can be followed by \email, \homepage, \thanks as well.
%\collaboration{}
%\noaffiliation

\date{\today}

\begin{abstract}
The realization of topological surface states and superconductivity within a single material platform is a crucial step toward achieving topologically nontrivial superconductivity. This can be achieved at an interface between a superconductor and a topological insulator, or within a single material that intrinsically hosts both superconductivity and topological surface states. Here we use scanning tunneling microscopy to study $\mathrm{Sn}_{1-x}\mathrm{In}_{x}\mathrm{Bi_{2}Te_{4}}$ crystals. Spectroscopic evidence reveals the coexistence of topological surface states and superconductivity on the same surface of the crystals. The Te-terminated surface exhibits a single U-shaped superconducting gap with a size of up to $311\;\mathrm{\mu eV}$, alongside Dirac bands outside the gap. Analysis of the vortex structure and differential conductance suggests weak-coupling s-wave superconductivity. The absence of observed zero modes suggests that shifting the Fermi level closer to the Dirac point of the topological bands is necessary to realize a topological superconducting state.
\end{abstract}

% insert suggested keywords - APS authors don't need to do this
%\keywords{Topological superconductivity, Topological surface states, Scanning tunneling microscopy}

%\maketitle must follow title, authors, abstract, and keywords
\maketitle

\section{\label{sec:IT}Introduction}

The interplay between superconductivity and nontrivial band topology offers a pathway to realizing exotic topological superconducting states, which manifest as Majorana fermionic mode that promises high-fidelity fault tolerant quantum computing. While interfacing superconducting and topological states in heterostructures via the superconducting proximity effect has been a common approach, the coexistence of superconductivity and topological states within a single crystalline material provides a more natural platform for realizing Majorana fermions, eliminating the challenges associated with complex interface conditions. Research efforts have thus focused on either inducing superconductivity in topological materials or uncovering topological states in superconductors. $\mathrm{SnBi_{2}Te_{4}}$, known as a topological insulator, has been shown to exhibit superconductivity under high pressure, accompanied by a structural transition \cite{Li2022}. Excitingly, recent work has demonstrated that alloying indium into $\mathrm{SnBi_{2}Te_{4}}$ induces bulk superconductivity, with critical temperatures ($T_{c}$) reaching up to $1.85\;\mathrm{K}$ at an indium content ($x$) of 0.61 \cite{McGuire2023}. This discovery raises fundamental interest in exploring the nature of superconductivity and its potential interaction with topological surface states in this material.

In this paper, we investigate the superconductivity and topological properties of $\mathrm{Sn}_{1-x}\mathrm{In}_{x}\mathrm{Bi_{2}Te_{4}}$ using scanning tunneling microscopy (STM). The superconducting gap in differential conductance ($dI/dV$) exhibits a BCS U-shaped single gap, indicating particle-hole symmetry. The size of the gap is $198\;\mathrm{\mu eV}$ for $x=0.33$ case and increases to $311\;\mathrm{\mu eV}$ when $x=0.60$. A hexagonal lattice of vortices is observed, suggesting the absence of strong vortex pinning sites. The flux per vortex matches the single magnetic flux quantum. The Ginzburg-Landau (GL) coherence length and GL parameter are obtained, and we compare the theoretical upper critical field ($H_{c2}$) with the experimental value. Linearly dispersive bands are observed with angle-resolved photoemission spectroscopy (ARPES), and the location of the Dirac point and its dependence on In doping are confirmed by STM. The Dirac point shifts downward from the Fermi level ($E_{F}$), reaching $-361\;\mathrm{meV}$ as the In doping ratio increases to $0.60$. This may account for the absence of zero-bias peaks in the vortex core. Our findings confirm the coexistence of superconducting and topological properties in van der Waals $\mathrm{Sn}_{1-x}\mathrm{In}_{x}\mathrm{Bi_{2}Te_{4}}$ crystals, calling for further experiments to tune the $E_{F}$ closer to the topological bands to explore potential topological superconductivity.

\section{\label{sec:MT}Methods}

Single crystals of $\mathrm{Sn}_{1-x}\mathrm{In}_{x}\mathrm{Bi_{2}Te_{4}}$ were synthesized following the procedure described in Ref. \cite{McGuire2023}. The indium content ($x$) in the crystals used for this study were $x=0.00$, $0.33$, and $0.60$, respectively.

Samples were measured using two STMs: a UNISOKU USM1600 operating at $4.2\;\mathrm{K}$ or $40\;\mathrm{mK}$ with 2-2-9 T vector magnets, and a SCIENTA OMICRON LT Nanoprobe operating at $4.6\;\mathrm{K}$. Platinum iridium (PtIr) tips were used for STM measurements. The samples were cleaved in ultrahigh vacuum (UHV) conditions for STM measurements, either at low temperatures (with cold cleaving stage at $83\;\mathrm{K}$ or within one minute after removing the sample from $4.6\;\mathrm{K}$ STM head) or at room temperature. After cleavage, the samples were cooled without magnetic field. STM topographic images were acquired in constant current mode, while $dI/dV$ was measured using a lock-in amplifier with a modulation voltage superimposed on the DC bias voltage. 

ARPES measurements were performed in a lab-based system with a SCIENTA DA30L analyzer and an $11\;\mathrm{eV}$ laser system. The base pressure is lower than $5\times 10^{-10}\;\mathrm{mbar}$ and the sample temperature is $\sim8\;\mathrm{K}$ during measurements. A pass energy $2\;\mathrm{eV}$ and $0.3\;\mathrm{mm}$ slit was used for a energy resolution $\sim3\;\mathrm{eV}$ and momentum resolution $\sim0.01\;\text{\r{A}}{}^{-1}$.

To evaluate the thermodynamic and electronic stability of the mixed phases in $\mathrm{Sn}_{1-x}\mathrm{In}_{x}\mathrm{Bi_{2}Te_{4}}$, we performed first-principles calculations based on density functional theory (DFT) using the VASP package \cite{Kresse1993,Kresse1996}. The projector augmented wave (PAW) pseudopotentials \cite{Blochl1994,Kresse1999} were employed to treat core-valence interactions, and Perdew-Burke-Ernzerhof (PBE) \cite{Perdew1996} for exchange-correlation functional. To investigate defect formation energetics and doping behavior in $\mathrm{SnBi_{2}Te_{4}}$, two supercell models were utilized: a $2\times2\times1$ supercell corresponding to a $25.0$\% concentration of point defects or dopants, and a $3\times3\times1$ supercell representing an $11.1$\% concentration. Both models were fully relaxed until the atomic forces were below $0.01\;\mathrm{eV}/\text{\r{A}}$, and the total energy was converged to within $10^{-6}\;\mathrm{eV}$. A kinetic energy cutoff of $500\;\mathrm{eV}$ was used for the plane-wave basis set. Brillouin zone integration was performed using $\Gamma$-centered Monkhorst-Pack grids, with a $3\times3\times1$ mesh for the $2\times2\times1$ supercell and a $2\times2\times1$ mesh for the larger $3\times3\times1$ supercell, ensuring consistent sampling density across configurations. All calculations were carried out without imposing symmetry constraints in order to capture local structural relaxations induced by defects and dopants.

\section{\label{sec:SS}Atomic structure of cleaved surfaces and stability analysis}

Single-crystal $\mathrm{SnBi_{2}Te_{4}}$ has a layered van der Waals structure, composed of three septuple layers within a unit cell, and crystallizes in the $R\overline{3}m$ space group (No. 166). This microstructure was confirmed by X-ray diffraction (XRD) and scanning transmission electron microscopy (STEM) measurements \cite{Li2022,Vilaplana2016,Saxena2023,Zou2018,Fragkos2021}. The 2D structure of each atomic layer, perpendicular to the lattice vector $\mathbf{c}$, is hexagonal. The lattice origin of each layer resides at one of three positions: $\mathbf{0}$, $(\mathbf{a}+2\mathbf{b})/3$, or $(2\mathbf{a}+\mathbf{b})/3$, where $\mathbf{a}$ and $\mathbf{b}$ are lattice vectors within a single layer. Ideally, each septuple layer follows a stacking sequence of Te-Bi-Te-Sn-Te-Bi-Te, with STM probing the outermost Te layers after cleavage.

\begin{figure}
\includegraphics[width=0.4\linewidth]{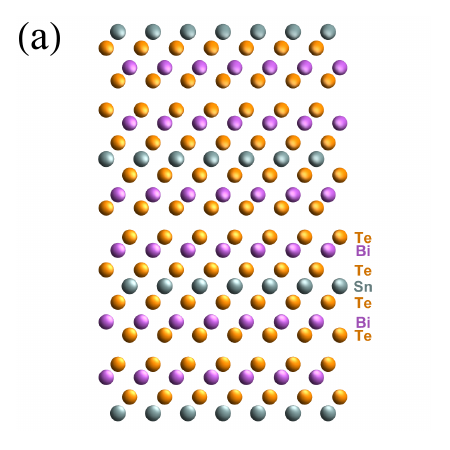}
\includegraphics[width=0.4\linewidth]{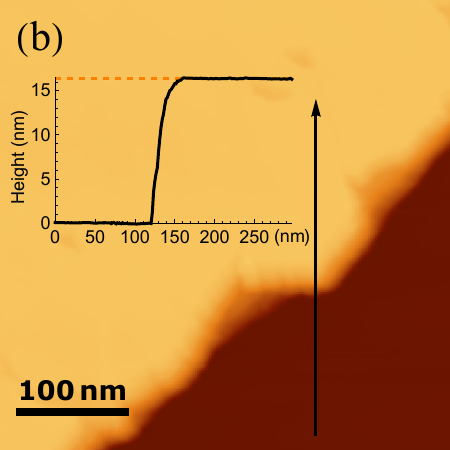}\\[0.5ex]
\includegraphics[width=0.4\linewidth]{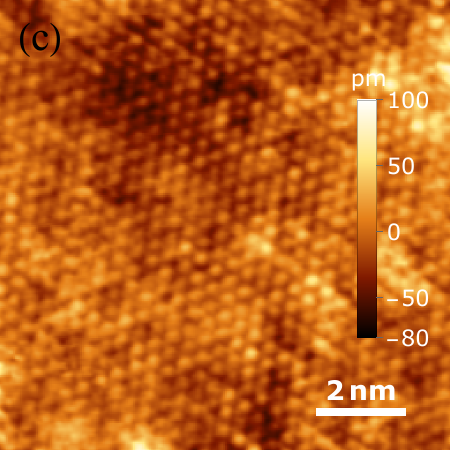}
\includegraphics[width=0.4\linewidth]{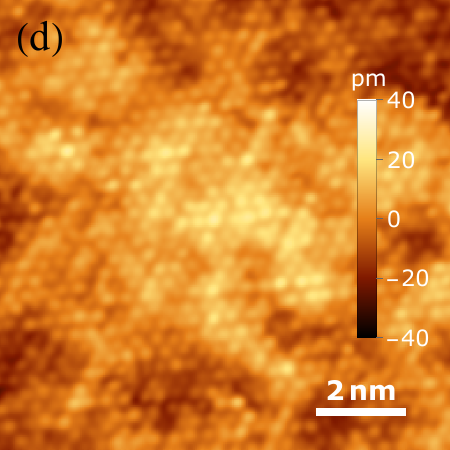}\\[0.5ex]
\includegraphics[width=0.4\linewidth]{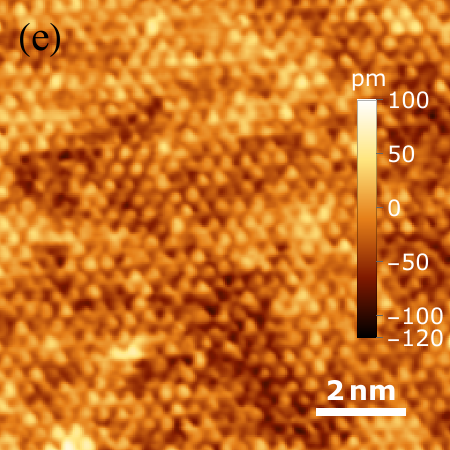}
\includegraphics[width=0.4\linewidth]{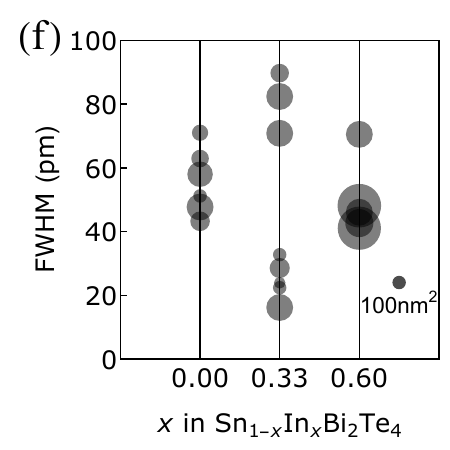}
\caption{\label{fig:SS}Crystal structure and STM topographic images of In doped $\mathrm{SnBi_{2}Te_{4}}$. (a) Crystal structure of $\mathrm{SnBi_{2}Te_{4}}$ illustrating the arrangement of Sn, Bi, and Te atoms as viewed along the $\mathbf{a}+\mathbf{b}$ direction, where $\mathbf{a}$ and $\mathbf{b}$ are lattice vectors. (b) A large-scale STM topographic image of Te-terminated terraces in $\mathrm{Sn_{0.4}In_{0.6}Bi_{2}Te_{4}}$ ($U=1\;\mathrm{V}$, $I=20\;\mathrm{pA}$). The inset shows a line profile along the arrow, indicating a step height of $16.4\;\mathrm{nm}$ between terraces, corresponding to $4$ unit cells or $12$ septuple layers. (c) Atomic-resolution STM topographic image of the Te layer in undoped $\mathrm{SnBi_{2}Te_{4}}$, cleaved at $83\;\mathrm{K}$ and measured at $40\;\mathrm{mK}$ ($U=10\;\mathrm{mV}$, $I=1\;\mathrm{nA}$). (d) Atomic-resolution STM topographic image of the Te layer in $\mathrm{Sn_{0.67}In_{0.33}Bi_{2}Te_{4}}$, cleaved at room temperature and measured at $4.3\;\mathrm{K}$ ($U=100\;\mathrm{mV}$, $I=1\;\mathrm{nA}$). (e) Atomic-resolution STM topographic image of the Te layer in $\mathrm{Sn_{0.4}In_{0.6}Bi_{2}Te_{4}}$, cleaved at room temperature and measured at $4.2\;\mathrm{K}$ ($U=100\;\mathrm{mV}$, $I=1\;\mathrm{nA}$). (f) Full width at half maximum (FWHM) of the height distribution calculated from atomic-resolution topographic images for three different In concentrations. All images were measured at the same setpoint ($U=100\;\mathrm{mV}$, $I=1\;\mathrm{nA}$). The FWHM was determined using histograms with a bin width of $1\;\mathrm{pm}$, and the size of each circle represents the area of the topographic image used for the FWHM calculation. A reference circle, corresponding to the FWHM for a $100\;\mathrm{nm}^{2}$ image, is shown at the bottom right of the plot.}
\end{figure}

Fig.~\ref{fig:SS} illustrates the atomic structure of a single-unit-cell height of $\mathrm{SnBi_{2}Te_{4}}$, STM topographic images of cleaved $\mathrm{Sn}_{1-x}\mathrm{In}_{x}\mathrm{Bi_{2}Te_{4}}$, and a full width at half maximum (FWHM) plot comparing height distributions for samples with different indium doping levels. The STM topographic images in Figs.~\ref{fig:SS}(c)-(e) show the hexagonal lattice of Te atoms, with dark and bright patches appearing across all samples, independent of the In concentration. These patches likely arise from inhomogeneity within the sublayer, which primarily consists of Bi atoms but also contains Sn atoms \cite{McGuire2023}. Previous studies have reported Bi-Sn cation exchange in single-crystal $\mathrm{SnBi_{2}Te_{4}}$, resulting in a reduction of the bulk gap \cite{Zou2018}. Similar Bi-Sn intermixing has been observed in molecular beam epitaxy grown films \cite{Fragkos2021}. However, whether In atoms substitute for Bi atoms upon doping remains uncertain. Stoichiometric analysis suggests that In predominantly replaces Sn atoms \cite{McGuire2023}.

To evaluate the effect of In doping on the electronic inhomogeneity of the Te surface, surface heights were binned from each STM topographic image, measured under identical setpoint ($U=100\;\mathrm{mV}$, $I=1\;\mathrm{nA}$) with a bin size of $1\;\mathrm{pm}$. The FWHM was then extracted and plotted, as shown in Fig.~\ref{fig:SS}(f). While the FWHM varies with the scanning area, no clear dependence on the In doping ratio was observed. This indicates that the doped In atoms do not significantly affect the outermost Te layer or the underlying Bi layer. This contrasts with a similar system $\mathrm{PbBi_{2}Te_{4}}$, where In atoms are known to substitute for both Pb and Bi atoms \cite{Xu2023}.

\begin{figure}
\includegraphics[width=0.435\linewidth]{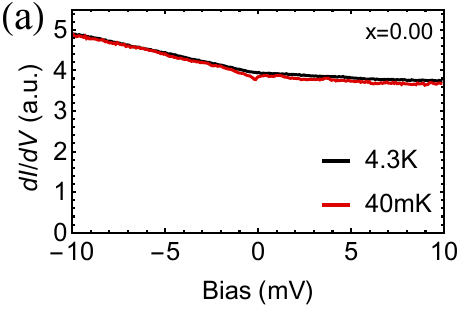}
\includegraphics[width=0.435\linewidth]{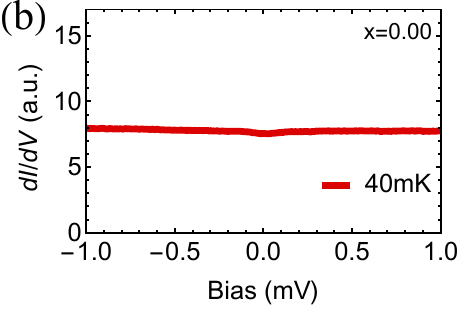}\\[0.5ex]
\includegraphics[width=0.435\linewidth]{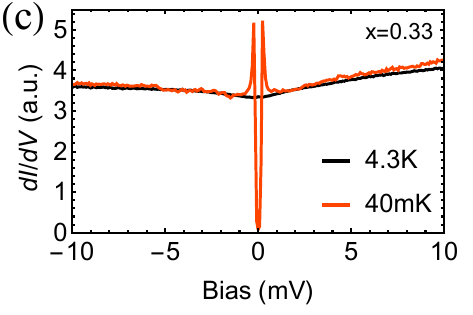}
\includegraphics[width=0.435\linewidth]{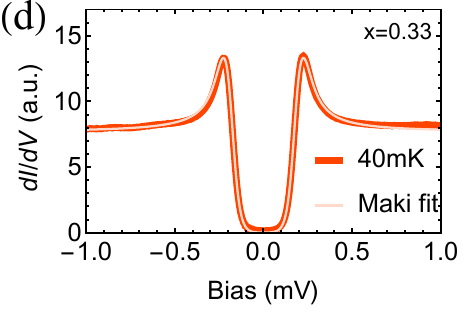}\\[0.5ex]
\includegraphics[width=0.435\linewidth]{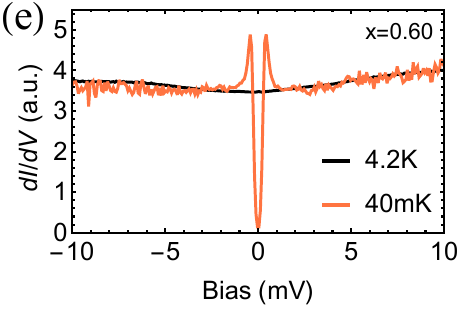}
\includegraphics[width=0.435\linewidth]{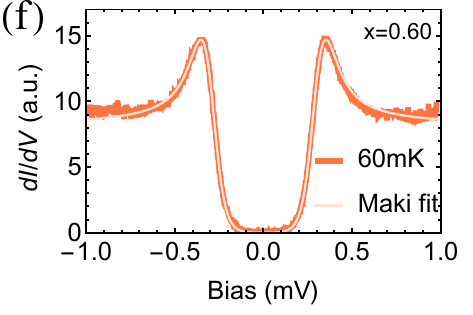}
\caption{\label{fig:GG:gap}$dI/dV$ spectra for $\mathrm{Sn}_{1-x}\mathrm{In}_{x}\mathrm{Bi_{2}Te_{4}}$ across different In concentrations ($x$) and temperatures. $dI/dV$ curves in (a), (c) and (e) were measured at two different temperatures to observe the superconducting transition ($U=10\;\mathrm{mV}$, $I=1\;\mathrm{nA}$, $\mathrm{mod}=50\;\mathrm{\mu V}$). $dI/dV$ curves in (b), (d) and (f) were measured at $40\;\mathrm{mK}$ or $60\;\mathrm{mK}$, with the setpoint $U=1\;\mathrm{mV}$, $I=1\;\mathrm{nA}$, $\mathrm{mod}=10\;\mathrm{\mu V}$. Curves in (d) and (f) show the superconducting gap and are fitted using the Maki function. Fitting parameters are ($\alpha=5.24\;\mathrm{\mu eV}$, $\Delta=198\;\mathrm{\mu eV}$, $T=244\;\mathrm{mK}$) for $dI/dV$ in (d), and ($\alpha=4.90\;\mathrm{\mu eV}$, $\Delta=311\;\mathrm{\mu eV}$, $T=378\;\mathrm{mK}$) for $dI/dV$ in (f).}
\end{figure}

The electronic structures, formation energies, and structural responses of In doped $\mathrm{SnBi_{2}Te_{4}}$ were systematically investigated using first-principles calculations for various defect and dopant configurations, including vacancies, intersite mixing, and indium substitution. To evaluate defect energetics at different concentrations, we constructed two supercells, $2\times2$ ($25.0$\%) and $3\times3$ ($11.1$\%) models. The formation energy $\Delta E_f$ of a vacancy or substitutional defect was calculated as\\
\[
\Delta E_f = E_{\text{defect}} - E_{\text{pristine}} + \sum_i n_i \mu_i,
\]where $E_{\text{defect}}$ and $E_{\text{pristine}}$ are the total energies of the defective and pristine supercells, respectively, and $n_{i}$ and $\mu_{i}$ represent the number and chemical potential of each atom added or removed. All vacancy types (Sn, Bi, $\mathrm{Te}_{in}$, and $\mathrm{Te}_{out}$) exhibited positive formation energies under Sn-, Bi-, and Te-rich conditions. The defect formation energies are greater than $+2.7\;\mathrm{eV}$ for Sn, $+3.8\;\mathrm{eV}$ for Bi, $+4.8\;\mathrm{eV}$ and $+4.4\;\mathrm{eV}$ for Te located inside or outside of the septuple layer, respectively, depending on the chemical potentials.  This indicates that the pristine $\mathrm{SnBi_{2}Te_{4}}$ structure is thermodynamically more stable than its defect-containing counterparts. Intersite mixing defects, such as atomic exchanges between Sn and Bi, Sn and Te, or Bi and Te, were also found to be energetically unfavorable, accompanied by only minor lattice distortions. In contrast, indium substitutional doping at the Sn ($\mathrm{In_{Sn}}$) and Bi ($\mathrm{In_{Bi}}$) sites yielded negative formation energies ($-1.0\;\mathrm{eV}$ and $-0.6\;\mathrm{eV}$, respectively) under In-, Te-, and Sn-rich conditions, indicating thermodynamic stability. These rich conditions correspond to experimental environments with an excess of elements available during synthesis. These results support experimental observations, which suggest that indium preferentially substitutes at the Sn site, although substitution at the Bi site is also feasible. In all cases, the associated lattice constant changes due to In doping were minimal ($<1$\%), indicating that the host lattice can accommodate indium with negligible structural disruption.

\begin{figure}
\includegraphics[width=0.32\linewidth]{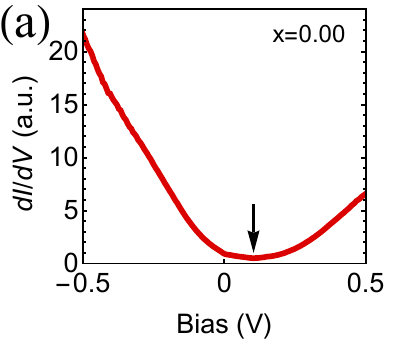}
\includegraphics[width=0.32\linewidth]{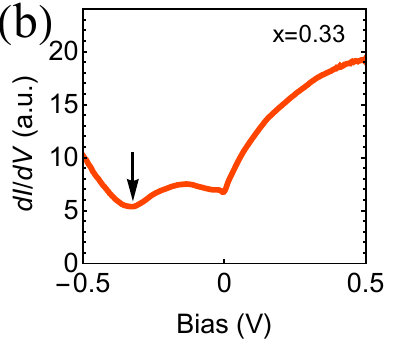}
\includegraphics[width=0.32\linewidth]{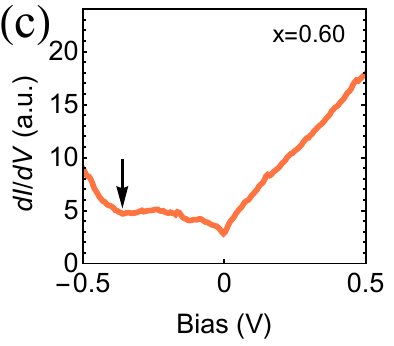}
\includegraphics[width=0.99\linewidth]{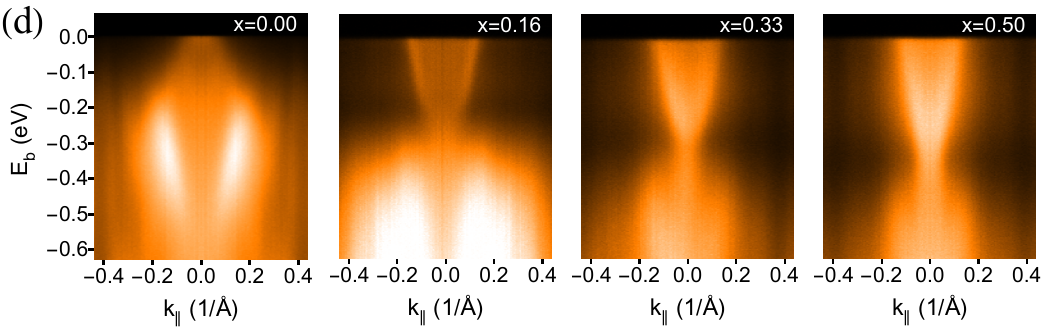}
\caption{\label{fig:GG:dirac}Dirac point detection for topological surface states of $\mathrm{Sn}_{1-x}\mathrm{In}_{x}\mathrm{Bi_{2}Te_{4}}$ using $dI/dV$ and ARPES. (a) Spatially averaged $dI/dV$ curve for the $x=0.00$ sample, measured over an $80\;\mathrm{nm}\times80\;\mathrm{nm}$ area ($32\times 32$ points, $U=800\;\mathrm{mV}$, $I=500\;\mathrm{pA}$, $\mathrm{mod}=10\;\mathrm{mV}$ at $973\;\mathrm{Hz}$). A local minimum is identified at $104\;\mathrm{meV}$. (b) Spatially averaged $dI/dV$ curve for the $x=0.33$ sample, measured over an $80\;\mathrm{nm}\times80\;\mathrm{nm}$ area ($32\times 32$ points, $U=500\;\mathrm{mV}$, $I=1\;\mathrm{nA}$, $\mathrm{mod}=10\;\mathrm{mV}$ at $973\;\mathrm{Hz}$). A local minimum is located at $-324\;\mathrm{meV}$. (c) Spatially averaged $dI/dV$ curve for the $x=0.60$ sample, measured over an $80\;\mathrm{nm}\times80\;\mathrm{nm}$ area ($32\times 32$ points, $U=800\;\mathrm{mV}$, $I=100\;\mathrm{pA}$, $\mathrm{mod}=10\;\mathrm{mV}$ at $1\;\mathrm{kHz}$). A local minimum is located at $-361\;\mathrm{meV}$. (d) ARPES data for four different In doping ratios ($x$). Dirac points from the linear dispersions are located at $50\;\mathrm{meV}$ ($x=0.00$), $-320\;\mathrm{meV}$ ($x=0.16$), $-350\;\mathrm{meV}$ ($x=0.33$), and $-380\;\mathrm{meV}$ ($x=0.50$).}
\end{figure}

\begin{figure*}
\includegraphics[width=0.15\linewidth]{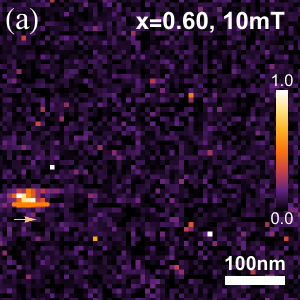}
\includegraphics[width=0.15\linewidth]{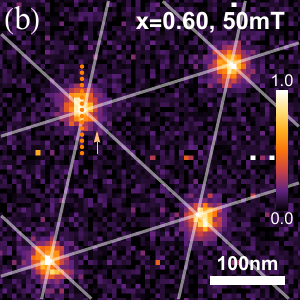}
\includegraphics[width=0.15\linewidth]{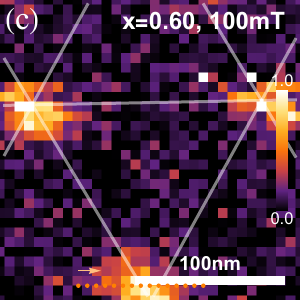}
\includegraphics[width=0.15\linewidth]{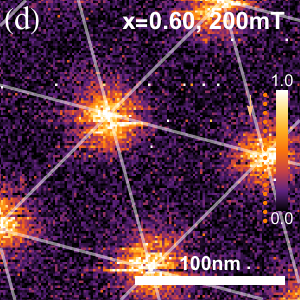}
\includegraphics[width=0.15\linewidth]{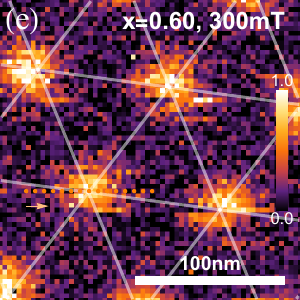}
\includegraphics[width=0.15\linewidth]{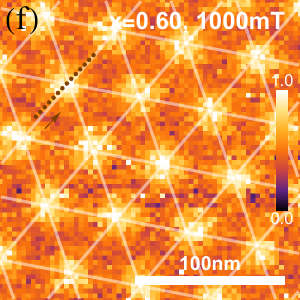}\\[1ex]
\includegraphics[width=0.15\linewidth]{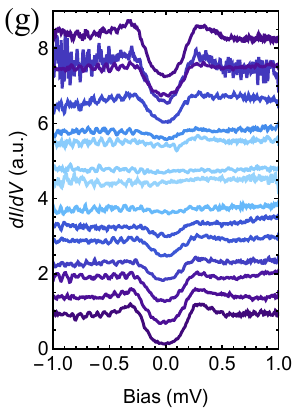}
\includegraphics[width=0.15\linewidth]{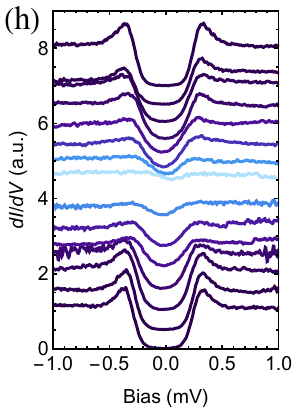}
\includegraphics[width=0.15\linewidth]{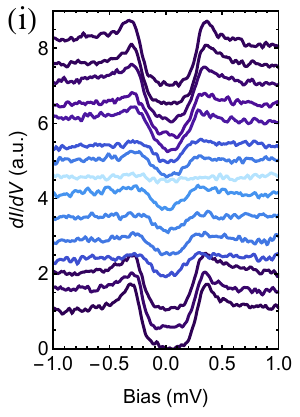}
\includegraphics[width=0.15\linewidth]{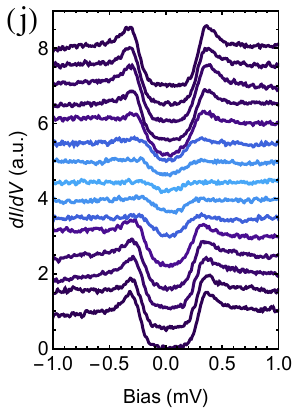}
\includegraphics[width=0.15\linewidth]{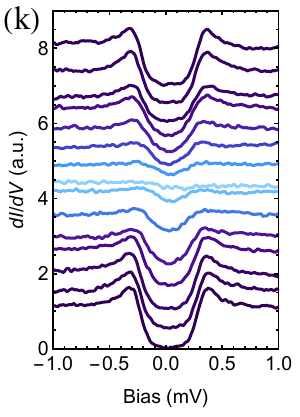}
\includegraphics[width=0.15\linewidth]{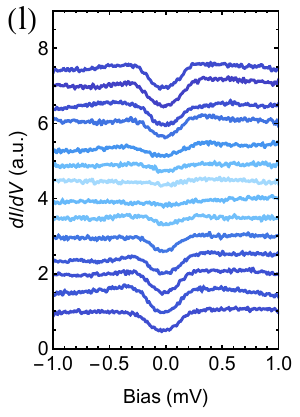}
\caption{\label{fig:VV:60}Vortex lattices at different vertical magnetic fields for the $x=0.60$ sample and the corresponding $dI/dV$ line spectra. (a)-(f) Normalized zero bias conductance maps showing the evolution of vortex lattices under varying vertical magnetic fields. White lines overlay the maps, representing the best-fit ideal hexagonal lattice structure. (g)-(l) Line $dI/dV$ spectra taken along lines passing through the vortex cores in the corresponding maps. The specific locations where $dI/dV$ measurements were taken are marked with dots.}
\end{figure*}

\section{\label{sec:GG}Superconducting gap and topological surface state}

Indium doping induces superconductivity in $\mathrm{SnBi_{2}Te_{4}}$, leading to a sharp transition to zero resistivity. The superconducting transition temperature ($T_{c}$) increases with the indium doping ratio $x$ in $\mathrm{Sn}_{1-x}\mathrm{In}_{x}\mathrm{Bi_{2}Te_{4}}$, reaching a maximum $T_{c}$ of $1.85\;\mathrm{K}$ at $x=0.61$ \cite{McGuire2023}. We measured the differential conductance $dI/dV$ of $\mathrm{Sn}_{1-x}\mathrm{In}_{x}\mathrm{Bi_{2}Te_{4}}$ before and after zero-field cooling from $4.2\;\mathrm{K}$ to $40\;\mathrm{mK}$, with the resulting spectra for three different In concentrations shown in Fig.~\ref{fig:GG:gap}. No superconducting gap was observed in the undoped ($x=0.00$) sample. However, superconducting gaps in the $dI/dV$ spectra were detected for $x=0.33$ and $x=0.60$ samples at $40\;\mathrm{mK}$ or $60\;\mathrm{mK}$. These gaps are spatially homogeneous and exhibit BCS-type characteristics.

The gaps were fitted using the Maki function \cite{Maki1964,Alexander1985,Worledge2000}, with the Fermi-Dirac distribution to account for broadening due to finite temperature and intrinsic noise. The fitting procedure yielded three parameters: effective temperature ($T$), orbital depairing parameter ($\alpha$), and superconducting gap at temperature $T$ ($\Delta(T)$). The $T_{c}$ was extracted from $T$ and $\Delta(T)$ using the relation $\Delta(0)=1.76k_{B}T_{c}$ and the approximate formula $\delta=\tanh(1.74\sqrt{\tau^{-1}-1})$ from BCS theory where $\delta=\Delta(T)/\Delta(0)$ is the normalized superconducting gap, $\tau=T/T_{c}$ is the reduced temperature, and $k_{B}$ is the Boltzmann constant. The calculated $T_{c}$ values from the two Maki fits of $dI/dV$ spectra in Figs.~\ref{fig:GG:gap}(d) and \ref{fig:GG:gap}(f) are $1.31\;\mathrm{K}$ for the $x=0.33$ sample and $2.05\;\mathrm{K}$ for the $x=0.60$ sample. These values are slightly higher than the transition temperatures of $1.22\;\mathrm{K}$ and $1.85\;\mathrm{K}$, obtained from resistivity or heat capacity measurements, suggesting a ratio $\Delta(0)/k_{B}T_{c}$ of $1.89$ or $1.95$, respectively, which is higher than the BCS value of $1.76$. This result is consistent with the heat capacity analysis \cite{McGuire2023}. Nevertheless, these values remain within the weak-coupling limit.

The spatially averaged $dI/dV$ and ARPES results shown in Fig.~\ref{fig:GG:dirac} indicate that the topological surface state persists even after In doping. The Dirac point in undoped $\mathrm{SnBi_{2}Te_{4}}$ lies above the Fermi level and the Dirac point shifts downward as the indium doping ratio $x$ increases.

\section{\label{sec:VV}Abrikosov vortex and flux quantum}

\begin{figure}
\includegraphics[width=0.35\linewidth]{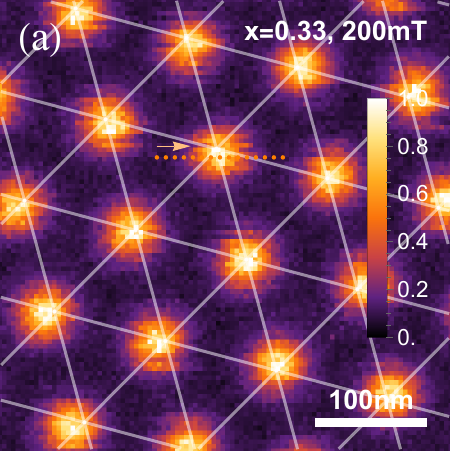}
\includegraphics[width=0.35\linewidth]{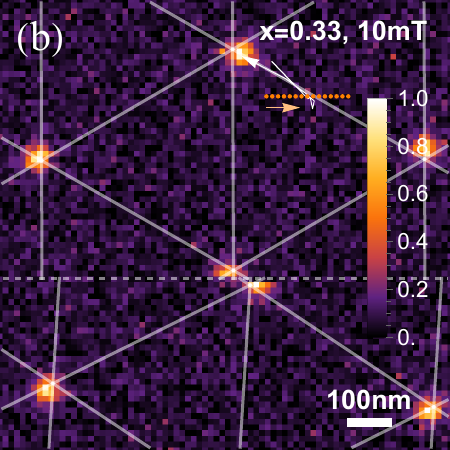}\\[1.0ex]
\includegraphics[width=0.35\linewidth]{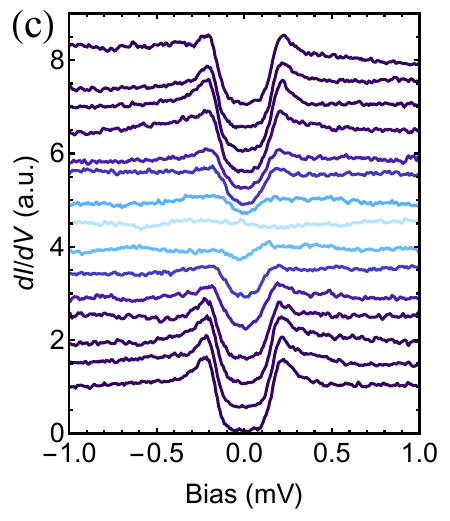}
\includegraphics[width=0.35\linewidth]{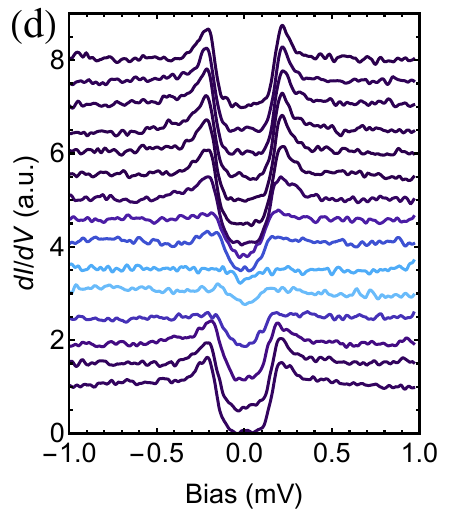}
\includegraphics[width=0.75\linewidth]{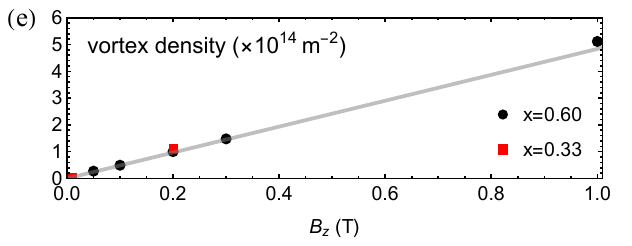}
\caption{\label{fig:VV:33}Vortex lattices at two different vertical magnetic fields for the $x=0.33$ sample along with the corresponding $dI/dV$ line spectra and vortex density plot. (a)-(b) Normalized zero bias conductance maps. The best-fit ideal hexagonal lattice structure is overlaid with white lines on each map. In (b), the dashed white line marks the row where the scanning tip changed the vortex lattice, and the white arrow highlights the trajectory of the vortex located at the top-middle part of the map during the line $dI/dV$ measurement. (c)-(d) Line $dI/dV$ spectra taken along paths passing through the vortex cores in the corresponding  nZBC maps. Each $dI/dV$ measurement point is indicated with a dot. (e) Vortex density as a function of vertical magnetic field, calculated from the best-fit hexagonal lattice constant for $x=0.60$ and $x=0.33$ samples. The gray line represents the theoretical density line, whose slope is $\Phi_{0}^{-1}$.}
\end{figure}

We confirmed type II superconductivity in indium doped $\mathrm{SnBi_{2}Te_{4}}$ by measuring zero bias conductance (ZBC) and observing a vortex lattice under external vertical magnetic fields in both $x=0.33$ and $x=0.60$ samples, as shown in Figs.~\ref{fig:VV:60} and \ref{fig:VV:33}.

Fig.~\ref{fig:VV:60} shows the magnetic field dependent vortex lattices and the $dI/dV$ spectra along lines passing through vortices for the $x=0.60$ sample. Since $dI/dV$ of this normal-insulator-superconductor (NIS) junction is proportional to the quasiparticle density of states, the $dI/dV$ spectrum at the center of a vortex is nearly flat, as the superconducting gap function approaches zero. To map the vortex lattice we used the normalized zero bias conductance (nZBC), defined as $dI/dV(V=0\;\mathrm{mV})$ divided by $dI/dV(V=1\;\mathrm{mV})$, which mostly falls within the range $(0,\;1)$. Figs.~\ref{fig:VV:60}(a)-(f) show maps of the nZBC at different magnetic fields, ranging from $10\;\mathrm{mT}$ to $1\;\mathrm{T}$. In the absence of a magnetic field, no features are observed in the nZBC. However, applying a vertical magnetic field induces a hexagonal vortex lattice. As shown in Figs.~\ref{fig:VV:60}(g)-(l), no in-gap states are detected around the vortices, and the $dI/dV$ spectra near the vortex centers become almost flat, indicating the complete suppression of the superconducting order parameter. At a vertical magnetic field of $1\;\mathrm{T}$, the nZBC does not fully diminish, even in the inter-vortex region, as illustrated in Figs.~\ref{fig:VV:60}(f) and \ref{fig:VV:60}(l). The hexagonal vortex lattice constant and its orientation vary with the vertical magnetic field. Each vortex position was used to compute the ideal hexagonal lattice vectors by minimizing the variance of distances from the centroid within a unit cell. The best-fit lattices are overlaid with white lines in Figs.~\ref{fig:VV:60}(b)–(f).

Fig.~\ref{fig:VV:33} shows the vortex lattices and the $dI/dV$ spectra through vortices for the $x=0.33$ sample. Similar to the $x=0.60$ sample, the lattice structure is hexagonal, and the $dI/dV$ spectra are nearly flat at the vortex cores. The dashed line in Fig.~\ref{fig:VV:33}(b) marks the row where the vortex lattice was altered by the STM tip. This unintended vortex manipulation frequently occurred at low magnetic fields due to weak vortex-vortex interactions. A demonstration of tip-induced vortex movement is indicated by a white arrow in Fig.~\ref{fig:VV:33}(b). The vortex lattice orientations have no preferred angle and can adopt multiple orientations depending on the initial conditions in any In doped samples, suggesting that the samples are homogeneous and free of pinning sites.

Each vortex carries a single magnetic flux quantum when the phase winding number of the superconducting order parameter is $1$. However, fractional or multi-quanta vortex states can emerge in multicomponent or multiband superconductors \cite{Tanaka2018,Iguchi2023,Gozlinski2023} or in finite-sized systems \cite{Tanaka2002,Kumar2015,VlaskoVlasov2023}. To calculate the magnetic flux per each vortex in In doped $\mathrm{SnBi_{2}Te_{4}}$, the vortex density (i.e., the number of vortices per square meter) was determined from the hexagonal vortex lattice constant for each applied magnetic field, as shown in Fig.~\ref{fig:VV:33}(e). Fitting these data points with a line passing through the origin yields a magnetic flux of $1.96\times 10^{-15}\;\mathrm{Wb}$ per vortex, assuming each vortex carries the same flux. This value closely matches the flux quantum predicted by Ginzburg-Landau theory, given by $\Phi_{0}=\frac{hc}{2e} = 2.07\times 10^{-15}\;\mathrm{Wb}$. The gray line in Fig.~\ref{fig:VV:33}(e) represents the theoretical density line, assuming each vortex carries a single magnetic flux quantum.

\section{\label{sec:GL}GL coherence length and $\mathbf{H_{c2}}$}

\begin{figure}
\includegraphics[width=0.83\linewidth]{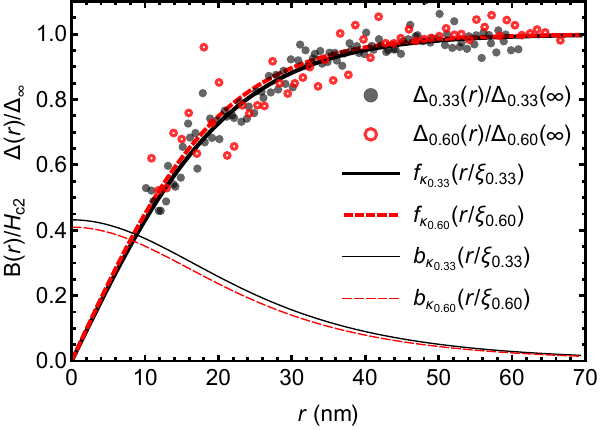}
\caption{\label{fig:GL:XI}Normalized gap $\Delta_{x}(r)/\Delta_{x}(\infty)$ as a function of distance $r$ from the center of the vortex in Figs. \ref{fig:VV:60}(b) and \ref{fig:VV:33}(a), normalized order parameter magnitude $f_{\kappa_{x}}(r/\xi_{x})$ fits to the normalized gap, and the corresponding magnetic field $b_{\kappa_{x}}(r/\xi_{x})$ in the units of $H_{c2}$ where $x$ represents the indium doping ratio, $b(r)=a'(r)+a(r)/r$, and $f(r)$ and $a(r)$ are solutions of Eqs.~(\ref{eq:GL:CYL1}) and (\ref{eq:GL:CYL2}). The GL parameters are $\kappa_{0.33}=1.29$, $\kappa_{0.60}=1.34$, and the GL coherence lengths are $\xi_{0.33}=15.9\;\mathrm{nm}$, $\xi_{0.60}=15.0\;\mathrm{nm}$, obtained from the fitting.}
\end{figure}

Coherence length $\xi$ and penetration depth $\lambda$ are two fundamental length scales in Ginzburg-Landau (GL) theory, governing the superconducting behavior of a material. The ratio of these two scales defines the GL parameter $\kappa=\lambda/\xi$, which determines whether a superconductor is type I ($\kappa<1/\sqrt{2}$) or type II ($\kappa>1/\sqrt{2}$). The penetration depth or its variation can be directly measured by techniques sensitive to magnetic fields, such as superconducting quantum interference device (SQUID) or tunnel diode oscillator \cite{Fletcher2007,Prozorov2011}. The coherence length can be extracted from the spatial variation of the superconducting order parameter, which can be measured by STM. Compared to microscopic theories, Ginzburg-Landau theory provides a more convenient approach for describing the spatial variation of the order parameter $\Psi(\mathbf{r})$ in the presence of a position-dependent magnetic vector potential $\mathbf{A}(\mathbf{r})$. However, GL theory is only valid under the assumption that the order parameter is small and varies slowly in space. The GL free energy $F_{\mathrm{GL}}=\int f_{\mathrm{GL}}\: d^{3}r$ is a functional of $\Psi$, $\Psi^{*}$, and $\mathbf{A}$ where
\begin{eqnarray}
f_{\mathrm{GL}}&=&\alpha |\Psi|^{2} + \frac{\beta}{2} |\Psi|^{4}\nonumber\\
&&+ \frac{1}{2m} \left| \left(\frac{\hbar}{i}\nabla -\frac{q}{c} \mathbf{A}\right)\Psi \right|^{2} + \frac{(\nabla \!\times\! \mathbf{A})^{2}}{8\pi}\label{eq:GL:FED}.
\end{eqnarray}
in Gaussian units and $\alpha = \alpha(T)$, $\beta = \beta(T)$ are real parameters that depend on temperature, and we assume that $\alpha$ and $\beta$ are position-independent. $m$ represents the mass of the Cooper pair, and $q=-2e$, where $e$ is the elementary charge. Taking functional derivatives of $F_{\mathrm{GL}}$ with respect to $\Psi^{*}$ and $\mathbf{A}$ leads to the following GL differential equations
\begin{eqnarray}
%\frac{\delta F_{\mathrm{GL}}}{\delta \Psi^{*}} =
\alpha \Psi + \beta |\Psi|^{2} \Psi +
\frac{1}{2m} \left( \frac{\hbar}{i}\nabla - \frac{q}{c} \mathbf{A} \right)^{2} \Psi =& \,0\label{eq:GL:GLDE1},\\
%\frac{\delta F_{\mathrm{GL}}}{\delta \mathbf{A}} =
-\frac{q\hbar}{2mci} \left( \Psi^{*}\nabla\Psi - \Psi\nabla\Psi^{*}\right) + \frac{q^{2}}{mc^{2}}|\Psi|^{2}\mathbf{A}\nonumber\\
+ \frac{\nabla \!\times\! (\nabla \!\times\! \mathbf{A})}{4\pi} = &\,0
\label{eq:GL:GLDE2}.
\end{eqnarray}
It is convenient to normalize $\Psi$, $\mathbf{A}$ and work with dimensionless variables by scaling $\mathbf{r}$. The scaling or normalization of variables can be done in different ways \cite{Zharkov2001,Oripov2020,Gropp1996,Peng2017}. We normalize the two GL differential equations using the relations $\Psi=|\Psi_{\infty}|\Psi'$, $\mathbf{A}=\sqrt{2}\lambda H_{c}\mathbf{A}'$, $\mathbf{r} = \xi \mathbf{r}'$, $\nabla = \frac{1}{\xi}\nabla'$ where $|\Psi_{\infty}|^{2} = -\frac{\alpha}{\beta}$,
%is the bulk Cooper pair density
$H_{c}$ is the thermodynamic critical field defined by $-\frac{H_{c}^{2}}{8\pi} = -\frac{\alpha^{2}}{2\beta}$, $\lambda(T) = \left( \frac{mc^{2}}{4\pi q^{2}|\Psi_{\infty}|^{2}} \right)^{1/2}$ is the London penetration depth, and $\xi(T) = \left( \frac{\hbar^{2}}{2m|\alpha(T)|} \right)^{1/2}$ is the GL coherence length.
$\mathbf{B}$ scales as $\mathbf{B}=\nabla \times \mathbf{A}= \sqrt{2}\kappa H_{c} \nabla' \times \mathbf{A}' = H_{c2}\mathbf{B}'$ and we obtain the two normalized GL differential equations after removing the prime symbols.
\begin{eqnarray}
&\Psi - |\Psi|^{2} \Psi + \left( \nabla - i\mathbf{A} \right)^{2} \Psi = 0\label{eq:GL:RGLDE1},\\
&\displaystyle\frac{i}{2} (\Psi^{*}\nabla\Psi - \Psi\nabla\Psi^{*})
+ |\Psi|^{2} \mathbf{A} + \kappa^{2} \nabla \!\times\! (\nabla \!\times\! \mathbf{A}) = 0\label{eq:GL:RGLDE2}.\quad
\end{eqnarray}
where $\kappa = \frac{\lambda(T)}{\xi(T)}$ is the GL parameter. If the order parameter varies only in the $x$-direction and the vector potential is ignored, we can obtain a real analytic solution of the form $\Psi(x)=\tanh(x/\sqrt{2})$ from the Eq.~(\ref{eq:GL:RGLDE1}) with the boundary conditions $\Psi(0)=0$ and $\Psi(\infty)=1$. Although this solution can be used to estimate the size of the vortex or the coherence length with zero bias tunneling conductance data \cite{Kohen2005,Ning2010,Song2013,Fente2016,Hoffmann2022,Eskildsen2002,Bergeal2006,Yin2009,Ding2023}, we solved Eqs.~(\ref{eq:GL:RGLDE1}) and (\ref{eq:GL:RGLDE2}) numerically without discarding the magnetic potential to extract $\kappa$ as well as $\xi$. Assuming that the order parameter does not vary along the $z$-direction, which is the same direction of the external magnetic field, and its magnitude has rotational symmetry about the vortex core axis, we can use cylindrical coordinates $(r,\theta,z)$. By substituting $\Psi=f(r)e^{in\theta}$ and $\mathbf{A}=a(r)\boldsymbol{\hat{\theta}}$ into the Eqs.~(\ref{eq:GL:RGLDE1}) and (\ref{eq:GL:RGLDE2}), we obtain the following equations.
\begin{eqnarray}
f''+\frac{f'}{r} - f \left( f^{2} -1 + \left( a-\frac{n}{r} \right)^{2} \right) = 0
\label{eq:GL:CYL1},\\
a''+\frac{a'}{r}-\frac{a}{r^{2}} - \frac{f^{2}}{\kappa^{2}} \left( a-\frac{n}{r} \right) = 0
\label{eq:GL:CYL2}.
\end{eqnarray}
where $n$ is the winding number. Here we set $n=1$ because we have already verified that each vortex carries a single magnetic flux quantum in the previous section. These coupled nonlinear GL differential equations can be solved simultaneously for given GL parameter $\kappa$. From Gor’kov derivation \cite{Gorgov1959}, GL order parameter $\Psi(r)$ is proportional to the BCS gap function $\Delta(r)$ in the region where the GL is valid, and thus we can fit the normalized gap data $\Delta(r)/\Delta(\infty)$ with the normalized order parameter $f_{\kappa}(r/\xi)$ as shown in the Fig.~\ref{fig:GL:XI} where $r$ is the distance from the center of a vortex. In the fitting process, we used $L_{1}$-norm to reduce the effect of outliers, and did not include gap data for $r<10\;\mathrm{nm}$ because magnetic field was not considered in the gap fitting with Maki function \cite{Maki1964} and the gap does not converge well to zero for nearly flat $dI/dV$ at the center of a vortex due to competing of gap fitting parameters. With $\kappa$ and $\xi$ obtained from the normalized gap fitting, $\lambda=\kappa \xi$, and $H_{c2}=\frac{\Phi_{0}}{2\pi\xi^{2}}$ from the GL theory. For $x=0.33$ sample, $\kappa=1.29$, $\xi=15.9\;\mathrm{nm}$ and $\lambda=20.5\;\mathrm{nm}$, $H_{c2}=13.0\;\mathrm{kOe}$. For $x=0.60$ sample, $\kappa=1.34$, $\xi=15.0\;\mathrm{nm}$ and $\lambda=20.1\;\mathrm{nm}$, $H_{c2}=14.6\;\mathrm{kOe}$. These $H_{c2}$ values are comparable to the estimated $H_{c2}(T=0)$ from experimental data, which are $10.3\;\mathrm{kOe}$ for $x=0.33$ and $13.8\;\mathrm{kOe}$ for $x=0.61$ \cite{McGuire2023}. We also note that $\int_{0}^{\infty} r b_{\kappa}(r) dr = 1$, confirming one vortex carries a single magnetic flux quantum $\Phi_{0}$.

% Data from vortices in Figs. (x=0.60, 50mT) and (x=0.33, 200mT)

\section{\label{sec:CC}Conclusions and Discussion}

We confirmed s-wave superconductivity by fitting the superconducting gap and comparing the calculated $T_{c}$ and $H_{c2}$ from STM data with experimental results from \cite{McGuire2023}. However, no in-gap states, including Majorana zero mode (MZM), and Caroli, de Gennes, and Matricon (CdGM) states \cite{Caroli1964} were observed near the vortex core in both the $x=0.33$ and $x=0.60$ systems. This could be attributed to the small energy spacing of CdGM states, $\Delta^{2}/E_{F}$, and the insufficient energy resolution of our measurements ($60\;\mathrm{\mu eV}$, assuming thermal broadening corresponding to an electron temperature of $\sim200\;\mathrm{mK}$). In contrast, CdGM states have been observed in the $\mathrm{FeTe_{0.55}Se_{0.45}}$ system \cite{Chen2018}, where a large $\Delta/E_{F}$ ratio was reported in ARPES measurements \cite{Rinott2017}, suggesting a BCS to BEC crossover regime. Most first-principles calculation studies of undoped $\mathrm{SnBi_{2}Te_{4}}$ show that the topological surface state and its Dirac point in the surface band structure overlap with the bulk valence band in energy, and the Dirac point is located below the Fermi level \cite{Dalui2023,Li2021,Menshchikova2020,Rongione2022}, even when Sn-Bi intermixing is considered \cite{Eremeev2023}. This is consistent with existing ARPES data, which confirm that the Dirac point is located below the Fermi level \cite{Li2021,Rongione2022,Eremeev2023}. However, our experimental results indicate that the Dirac point is above the Fermi level in undoped $\mathrm{SnBi_{2}Te_{4}}$ and below the Fermi level after In doping as shown in Fig.~\ref{fig:GG:dirac}, suggesting a bulk band contribution near the Fermi level, regardless of In doping.

There is a report highlighting the importance of the position of the Dirac point for the TI-SC $\mathrm{Bi_{2}Te_{3}}$/$\mathrm{NbSe_{2}}$ heterostructure system, where nontrivial ZBC was observed only when the thickness of the TI reached 5--6 quintuple layers (QLs), even though the topological surface state band formed at 3 QL. This behavior is attributed to the Dirac point being farther from the Fermi level in the 3 QL sample \cite{Xu2015}. Furthermore, MZMs may disappear due to the bulk band effects resulting from chemical potential tuning \cite{Hosur2011}. The shift of the Dirac point to negative energies relative to the Fermi level must be considered when discussing the role of In dopants, as In does not inject holes into the system as initially expected based on its valence electron count.

There have been numerous efforts to modify the electronic structure of undoped $\mathrm{SnBi_{2}Te_{4}}$ and related compound $\mathrm{PbBi_{2}Te_{4}}$. For example, the electronic structure of $\mathrm{SnBi_{2}Te_{4}}$ has been shown to depend on pressure \cite{Vilaplana2016} and temperature \cite{Dalui2023}. Increasing the Bi concentration leads to a new van der Waals structure consisting of $\mathrm{SnBi_{2}Te_{4}}$ septuple layers and $\mathrm{Bi_{2}Te_{3}}$ quintuple layers \cite{Fragkos2021}. In the topological insulator $\mathrm{PbBi_{2}(Te,Se)_{4}}$, added Se atoms prefer to occupy two inner layers in a septuple layer (SL) \cite{Hattori2020}, and a superconducting transition has been observed in In doped $\mathrm{PbBi_{2}Te_{4}}$ \cite{Xu2023}. Further tuning of In-doped $\mathrm{SnBi_{2}Te_{4}}$ system, while maintaining both superconductivity and topological surface state, is needed to facilitate the emergence of MZM and to gain a deeper understanding of its electronic properties.\\

\section{\label{sec:AK}Acknowledgments}

This work was supported by the U.S. DOE, Office of Science, National Quantum Information Science Research Centers, Quantum Science Center, and by the U.S. Department of Energy, Office of Science, Office of Basic Energy Sciences, Materials Sciences and Engineering Division. The STM measurement was conducted at the Center for Nanophase Materials Sciences (CNMS), which is a U.S. Department of Energy, Office of Science User Facility. This research used resources of the Oak Ridge Leadership Computing Facility at the Oak Ridge National Laboratory, which is supported by the Office of Science of the U.S. Department of Energy under Contract No. DE-AC05-00OR22725.

\bibliography{main}

\end{document}